# Electromagnon resonance in a collinear spin state of a polar antiferromagnet $Fe_2Mo_3O_8$

T. Kurumaji[1], Y. Takahashi[1, 2, 3], J. Fujioka[2], R. Masuda[2], H. Shishikura[2], S. Ishiwata[2, 3], and Y. Tokura[1, 2]

[1]RIKEN Center for Emergent Matter Science (CEMS), Wako 351-0198, Japan.
[2]Department of Applied Physics and Quantum Phase Electronics Center (QPEC), University of Tokyo, Tokyo 11-8656, Japan.
[3]PRESTO, Japan Science and Technology Agency, Chiyoda, Tokyo 102-8666, Japan

**Abstract**

Magnetic excitations are investigated for a hexagonal polar magnet $Fe_2Mo_3O_8$ by terahertz spectroscopy. We observed magnon modes including an electric-field active magnon, electromagnon, in the collinear antiferromagnetic phase with spins parallel to the $c$ axis. We unravel the nature of these excitations by investigating the correlation between the evolution of the mode profile and the magnetic transition from antiferromagnetic to ferrimagnetic order induced by magnetic field or Zn-doping. We propose that the observed electromagnon mode involves the collective precession of the spins with oscillating in-plane electric polarization through the mechanism of the linear magnetoelectric effect.

Cross correlation between magnetism and electricity, i.e., magnetoelectric (ME) effect, is a key in designing electric-field-controllable spin devices [1, 2]. Among various ME materials [3], multiferroics, which exhibit simultaneous magnetic and ferroelectric orders, have attracted tremendous interest because of recent discoveries of strong ME response upon magnetic phase transition as in $TbMnO_3$ [4] as well as of room-temperature multiferroics such as $BiFeO_3$ [5] and hexaferrite [6]. Entanglement between magnetism and electricity can be extended to elementary excitations, which was theoretically discussed since 1970's [7]. In fact, the electric-dipole-active magnon, termed the electromagnon, was observed as an infrared absorption in terahertz region [8]. Such excitations have been identified in various multiferroic materials [9], and promise new terahertz functionalities of multiferroics including optical control of magnetism and nonreciprocal directional dichroism [10, 11].

According to Khomskii [12], multiferroics can be classified into two types; in type-I multiferroics, the ferroelectricity and the magnetism have distinct origins, while the magnetic order itself is the driving force of ferroelectricity in type-II multiferroics. The former group includes $BiFeO_3$ and hexagonal $YMnO_3$ [5, 13], which show relatively large spontaneous polarizations and high ferroelectric/magnetic transition temperatures, while the magnetism only modestly influences the polarization and/or dielectric constant. The latter group including orthorhombic (perovskite type) $RMnO_3$ ($R$: rare earth) [4, 14], $Ni_3V_2O_8$ [15], and $MnWO_4$ [16] hosts strong ME coupling, while tending to show relatively lower transition temperatures partly because of the frustration in spin interactions.

Early works on type-II multiferroics including $R$MnO$_3$ [8], hexaferrites [17], and CuO [18] have clarified that incommensurate spiral magnetic orders generally exhibit the electromagnon resonances. Therein, a part of electromagnons is driven by the exchange striction mechanism described by the inner product of spins, i.e., $\bm{S}_i \cdot \bm{S}_j$ [19], while the inverse Dzyaloshinskii-Moriya (DM) mechanism expressed by $\bm{S}_i \times \bm{S}_j$ also contributes to the electromagnon resonance with both electric- and magnetic-dipole activities [11]. For type-I multiferroics, on the other hand, the electromagnons, which tend to show up less conspicuously in the spectra, are often connected to complex magnetic structures such as cycloidal magnetic order in BiFeO$_3$ [20, 21], and noncollinear multi-sublattice ferrimagnetic order in CaBaCo$_4$O$_7$ [22]. Here we report one other type of electromagnon in a type-I multiferroic Fe$_2$Mo$_3$O$_8$ with a simple collinear magnetic order of magnetic moments of Fe$^{2+}$ ions. We identify the mode characters of magnetic excitations including electromagnon by the polarization selection rule and the comparison with ferrimagnetic phase induced by chemical doping as well as by the magnetic field. We propose a model of magnetic excitations, where the ME coupling is taken into account, that consistently explains magnetic excitation in antiferromagnetic and ferrimagnetic phases.

Fe$_2$Mo$_3$O$_8$ forms a hexagonal lattice belonging to a polar space group

$P6_3mc$ (Fig. 1(a)). There exist two types of magnetic site for $Fe^{2+}$ ion, $A$ and $B$, characterized by the tetrahedral and octahedral coordination of oxygen, respectively [23]. $AO_4$ and $BO_6$ polyhedra share their corners to form a honeycomb lattice in the $ab$-plane. The two magnetic layers, which are related with a nonsymmorphic operation based on the $c$-glide plane with each other, are involved in the unit cell. Below the Néel point ($T_N$ = 60 K) the system evolves into the collinear antiferromagnetic (AF) state (see the inset of Fig. 1(b)) [24]. Application of magnetic field ($H_{dc}$) along the $c$ axis induces a collinear ferrimagnetic (FM) order [25, 26] (the inset of Fig. 1(c)). Alternatively, the FM state is stabilized also by substitution of more than 12.5 % of Fe with Zn [24, 26, 27]. Coexistence of spontaneous polarization and magnetic order below the transition temperature allows strong ME coupling and large linear ME coefficients in both the in-plane and out-of-plane components, which promises the characteristic spin wave excitation responding to ac electric/magnetic field of light.

Single crystals of $Fe_2Mo_3O_8$ and $(Zn_{0.125}Fe_{0.875})_2Mo_3O_8$ were grown by chemical vapor transport reaction as described in Refs. [28, 29] from the stoichiometric mixture of $MoO_2$, Fe, $Fe_2O_3$, and ZnO. Samples with $ab$-plane and $ac$-plane cut, whose dimensions are typically 2 x 2 mm², were prepared. The time-domain terahertz spectroscopy was employed to measure the refractive indices in a frequency range of 0.5 - 2.8 THz and the details about the experimental setup and procedures are described in Ref. [30]. Laser pulses with 100-fs duration from a Ti: sapphire laser were split into two paths to generate and detect the wave form of terahertz pulses. A ZnTe (110) crystal and a dipole antenna were used for generation and detection of terahertz pulses, respectively. The $H_{dc}$ was applied to the sample with a superconducting magnet in Voigt geometry, i.e., a light propagation vector $k^\omega$ perpendicular to $H_{dc}$.

Figures 1(d)-(f) show the spectra of extinction coefficient $\kappa$ (imaginary part of refractive index) for $Fe_2Mo_3O_8$ in zero field at 4.5 K for three possible geometries. As shown in Fig. 1(d), two clear resonance peaks are observed around 1.2 THz and 2.7 THz for the light polarized $E^\omega \perp c$ and $H^\omega \perp c$, denoted as EM and MM1, respectively. The characters of magnetic excitations can be deduced by the polarization selection rule derived from the results in Figs. 1(e)-(f); EM is concluded as electric-dipole (E1) active, i.e., electromagnon, because it can be excited by the $E^\omega \perp c$ (Fig. 1(e)) but not by $H^\omega \perp c$ (Fig. 1(f)), while MM1 is active for $H^\omega \perp c$ (not with $E^\omega \perp c$), indicating its magnetic-dipole (M1) active nature. To check the correlation between the mode profile and the magnetic order, we also measured the spectra for the collinear ferrimagnetic phase in the doped sample ($y = 0.125$) (Figs. 1(g)-(i)). This composition shows the FM state even at zero field (Fig. 1(c)). A single resonance peak is observed around 2.6 THz (MM2) (Fig. 1(g) and 1(i)), while no discernible resonance structure is seen around 1.2 THz. Thus, the electromagnon resonance is absent (Fig. 1(h)) in the current energy window, while the MM2 is active for $H^\omega \perp c$ (Fig. 1(i)) similarly to the MM1.

Figures 2(a)-(h) show the spectra of $\kappa$ for $y = 0$ and $y = 0.125$ at selected temperatures. With increasing temperature, the absorption of each mode gradually wanes and disappears above the transition temperature (Fig. 2 (d) and (h)). The temperature dependence of the spectral weights ($\propto -\frac{1}{d}\int \ln(t - t_0)\,d\omega$) for the respective modes are shown in Figs. 2(i) and 2(j). Here, $d$ is the thickness of the sample, $t$ the transmittance, and $\omega$ the angular frequency of light. $t_0$ is assumed to be a background due to flat absorption. We define the spectral weights of EM, MM1 and MM2 by the integration between 1.1 ~ 1.4 THz, 2.5 ~ 2.8 THz, and 2.4 ~ 2.6 THz, respectively. The magnitude of these resonances start to rise upon the magnetic ordering as shown in Figs. 2(i) and 2(j). This result indicates that the observed modes are collective excitations arising from the magnetic ordering and not from a gap excitations related to the crystal field. Indeed, the latter excitation was observed in noncentrosymmetric $Ba_2CoGe_2O_7$ [31, 32], which has E1 activity but is observable even above the transition temperature unlike the present case.

To clarify the mode characters in AF and FM states, the $H_{dc}$ dependence of magnetic resonances are measured at selected temperatures as summarized in Fig. 3. Figure 3(a) shows $\kappa$ spectra for AF state ($y = 0$, see the phase diagram in Fig. 1(b)) at 4.5 K under $H_{dc}//c$ for two different light polarizations. Figure 3(d) shows field evolution of excitation frequency. EM shows little magnetic field dependence, while the MM1 splits into two modes, implying the character of the conventional antiferromagnetic resonance. Although the EM may be also doubly degenerate, the possible frequency splitting appears to be too small to be detected for $\mu_0 H_{dc}$ up to 7 T.

We also performed the comparative measurements for the FM state stabilized by the magnetic field at 50 K for $y = 0$ and by the chemical doping at 4.5 K for $y = 0.125$. The data set are displayed in Figs. 3(b), 3(c), 3(e), and 3(f). At 50 K, $Fe_2Mo_3O_8$ shows the metamagnetic transition at $\mu_0 H_{dc} \sim 5.2$ T as shown by $M$-$H_{dc}$ curve in Fig. 3(e). Upon the transition, the EM in $E^\omega \perp c$ geometry suddenly disappears, while the split branches of the MM1 turn into a single mode (termed here MM2) with a slightly lower frequency (see the spectra for $\mu_0 H_{dc}= 5.1$ T and 5.3 T with $H^\omega \perp c$ in Fig. 3(b) and 3(e)). The similar MM2 mode in the FM phase are also exemplified by the FM phase induced by the chemical doping ($y = 0.125$), in which the monotonous softening of the MM2 is observed as the magnetic field is increased from 0 T to 7 T (Fig. 3(c) and 3(f)).

The emergence of the electromagnon mode in the AF phase indicates that the magnetic excitation possesses in-plane oscillation of electric polarization. The linear ME effect at the DC limit observed in FM phase [26] can be related to the electrical activity of magnon excitation in AF phase. Here, we consider two mechanisms for the linear ME effect as identified in Ref. [26], i.e., the inverse DM effect and the single-site anisotropy effect. Although the conventional inverse DM model in Refs. [33, 34] predicts $P$ along the $c$ ($z$) axis for the adjacent spin on $A$ and $B$ site, the local site asymmetry in the present compound allows $P$ in general directions, in accord with descriptions in Refs.

[35, 36]. In the case of the nearest-neighboring $A$ and $B$ sites in a honeycomb layer shown in Fig. 4(a), the symmetry with respect to the $zy$ plane allows in-plane electric polarization $p_y$ proportional to dynamical $x$ component of $\boldsymbol{S}_A \times \boldsymbol{S}_B$, i.e., $\delta S_{Ay}S_{Bz} - S_{Az}\delta S_{By}$. On the other hand, the single-site anisotropy effect [37, 38] induces the in-plane polarization $p_{iy}$ at each $i$-th Fe site due to canting of a spin as $p_{iy} \propto S_{iz}\delta S_{iy}$.

Here we speculate possible modes of magnon excitations to explain the observed resonance including EM, MM1 and MM2. We ignore, to the first approximation, the interlayer magnetic interactions, because the stacking honeycomb layers are intervened by a Mo layer. In each magnetic layer, spins at neighboring $A$ and $B$ sites prefer to oscillate in an antiferromagnetic manner as shown schematically in Fig. 4(a). The neighboring spins cant with slightly different angles into opposite direction, as the result of single-ion anisotropy at each site. In this circumstance, electric polarization due to inverse DM effect ($p_y$) is nonzero since the dynamical $x$ component of $\boldsymbol{S}_A \times \boldsymbol{S}_B$ ($\delta S_{Ay}S_{Bz} - S_{Az}\delta S_{By}$) is nonzero, and the difference of the transverse component of the respective spins ($\delta S_{Ay}$ and $\delta S_{By}$ in Figs. 4(a)) induces a net magnetization $m_y$. Note that the mutual relation between $m_y$ and $p_y$ is opposite for the upper and bottom layers (Fig. 4(a)) in the unit cell, since their relative positions are interchanged. Next, we take into account the interlayer coupling. In that case, doubly degenerate modes for the upper and bottom layers are coupled in in-phase or out-of-phase manner, resulting in the mode splitting. Figures 4(b) and 4(c) show in-phase and out-of-phase oscillations, respectively; $m_y$ ($m_1$ and $m_2$) and $p_y$ ($p_1$ and $p_2$) are shown for each layer. The oscillation pattern in Fig. 4(b) induces net $p_y$ while the $m_y$ cancels; this explains why the EM can be excited by the in-plane electric field but not by the in-plane magnetic field of light. As for the out-of-phase oscillation (Fig. 4(c)), $m_y$ remains finite while $p_y$ is cancelled; this corresponds to the MM1. Therefore, the configurations shown in Figs. 4(b) and 4(c) qualitatively explain the selection rule for the electromagnon and magnon modes observed in the AF phase. In Ref. [24], the interlayer coupling energies were estimated by the molecular field theory, i.e., the antiferromagnetic interlayer coupling between $A$ sublattices is stronger (~57 K) than that between $A$ and $B$ sublattices (~38 K). This is consistent with the lower excitation energy of EM than that of MM1; EM keeps the antiferromagnetic nature between interlayer $A$ sites during the oscillation as shown in Fig. 4(b), while MM1 violates it (Fig. 4(c)). Note that the in-plane electric polarizations due to the single-site anisotropy effect are uncancelled and cancelled for the spin configurations in Figs. 4(b) and 4(c), respectively, giving the same conclusion on the dipole-activities.

The above scheme is also applicable to the FM phase, which suggests a both E1 and M1 active mode (Fig. 4(d)) as well as a silent mode (mode X as shown in Fig. 4(e)), although the experimentally observed MM2 appears to be M1 active but least E1 active. From the symmetry point of view, four magnetic excitation branches exist for a four-sublattice collinear magnetic system. Thus, we believe there is another higher energy mode out of the range of this experiment which would show strong E1 and weak M1 activities, complementary to the nature of the MM2.

In conclusion, we observe two distinct collective magnetic excitations driven by electric and magnetic field of terahertz light, respectively, in the antiferromagnetic phase for polar magnet $Fe_2Mo_3O_8$. We have also revealed distinct properties of magnetic excitations for the antiferromagnetic and ferrimagnetic phases. The origin of the observed electromagnon is accounted for by the oscillation of electric polarization induced by precession of spins through the inverse Dzyaloshinskii-Moriya interaction and/or single-site anisotropy. Possible spin configuration for the excitations are suggested, which remains electric polarization uncanceled because of the out-of-phase interlayer coupling. The present observations show that the simple collinear magnetic order in type-I multiferroics can host electromagnon modes, promising versatile optical magnetoelectric phenomena in terahertz region as well as those in type-II multiferroics.

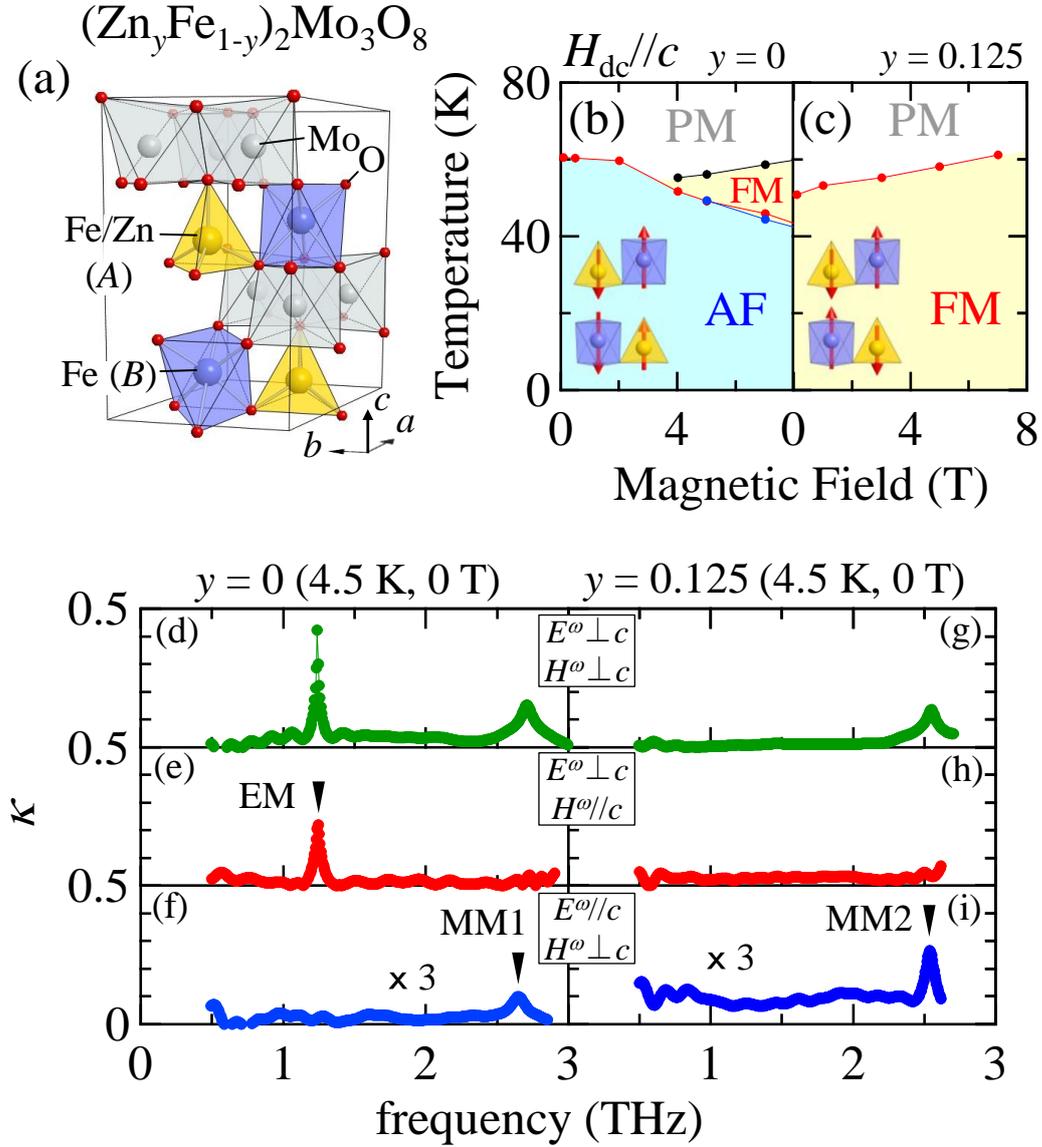

Fig. 1

(a) Crystal structure of $(Zn_yFe_{1-y})_2Mo_3O_8$. (b)-(c) Magnetic field ($H_{dc}$) vs. temperature phase diagrams under $H_{dc}//c$ for $y = 0$ and $y = 0.125$, respectively, as reproduced from Ref. [26]. Magnetic structure of each phase is also shown. (d)-(i) Spectra of $\kappa$ (imaginary part of refractive index, i.e., extinction coefficient) for respective light polarizations at 4.5 K in zero field for $y = 0$ ((d)-(f)) and for $y = 0.125$ ((g)-(i)).

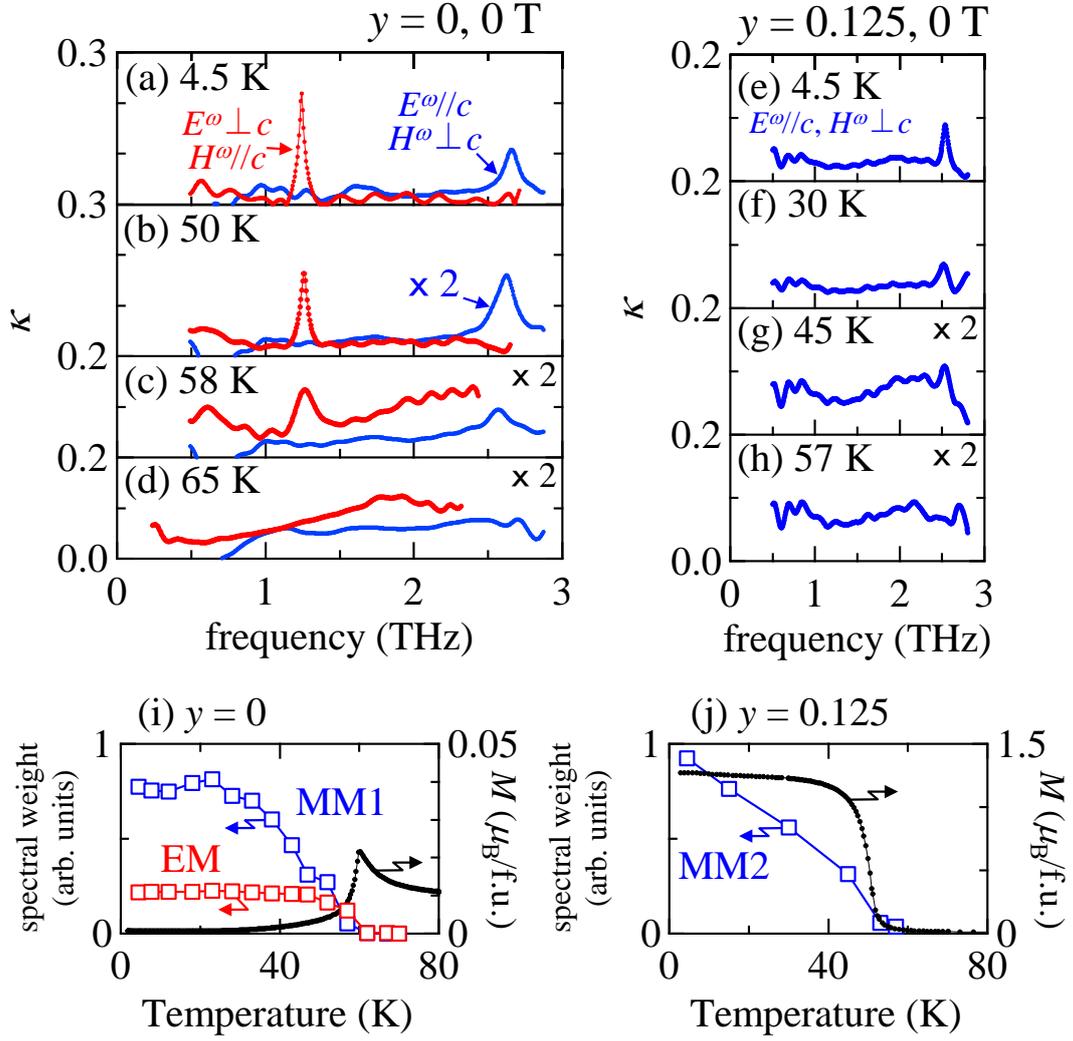

Fig. 2

(a)-(d) Temperature dependence of $\kappa$ for $y = 0$ in zero field. Red (blue) curves are for EM (MM1) measured with different light polarizations. (e)-(h) Corresponding spectra of $y = 0.125$ for MM2. Temperature dependence of spectral weight for (i) EM and MM1, and (j) MM2 in zero field. Magnetization measured with $\mu_0 H_{dc} = 0.1$ T is also shown for comparison.

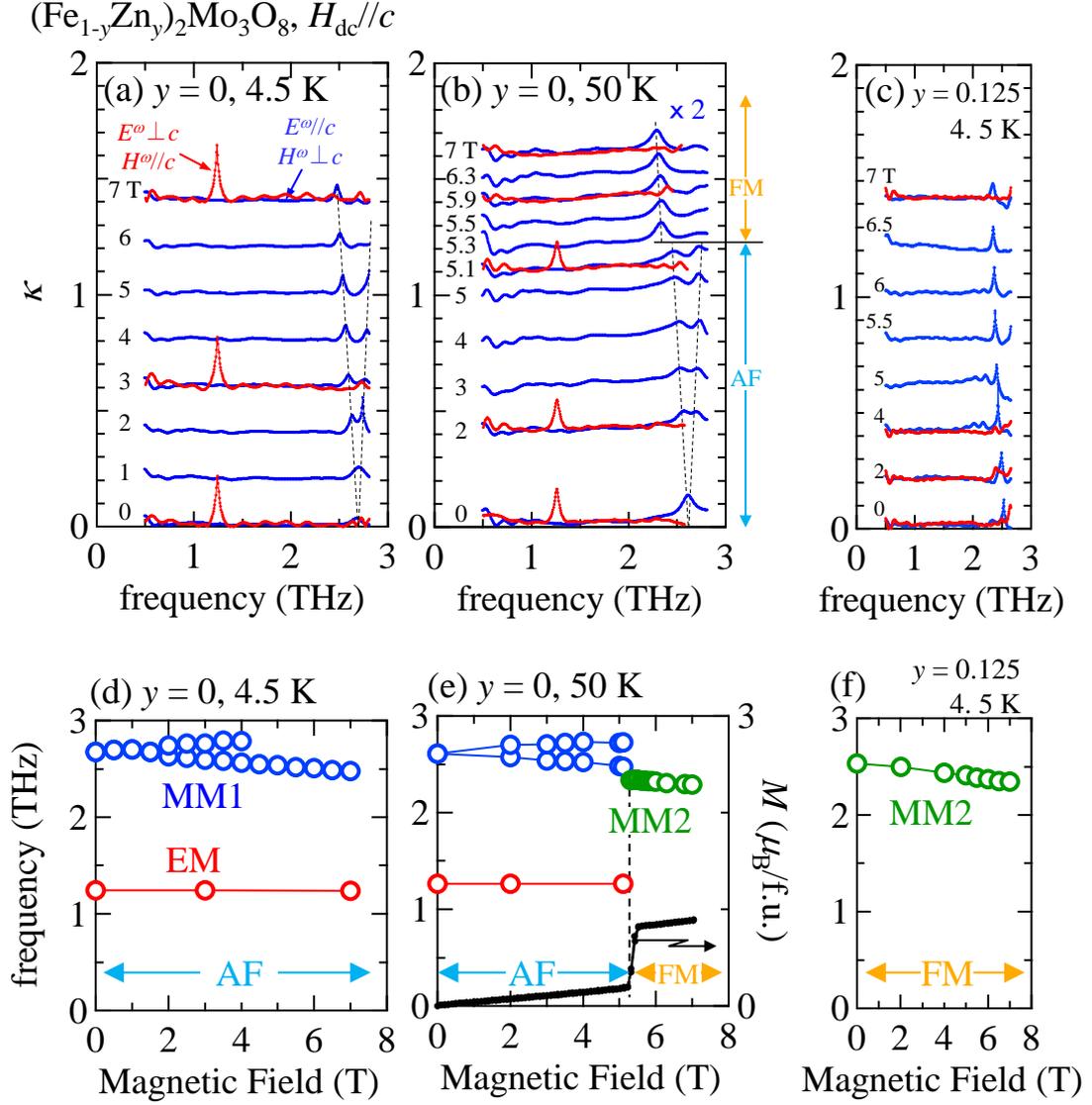

Fig. 3

(a)-(c) $\kappa$ spectra under various field magnitudes. Data are shifted vertically for clarity. Red (blue) curves are for the polarization $E^\omega \perp c$ and $H^\omega // c$ ($H^\omega \perp c$ and $E^\omega // c$). Blue curves in (b) are magnified by two. (d)-(f) Evolution of excitation frequency with $H_{dc}//c$: red, blue, and green circles are for EM, MM1, and MM2, respectively. In (e), magnetization along the $c$ axis at 50 K is also shown for comparison.

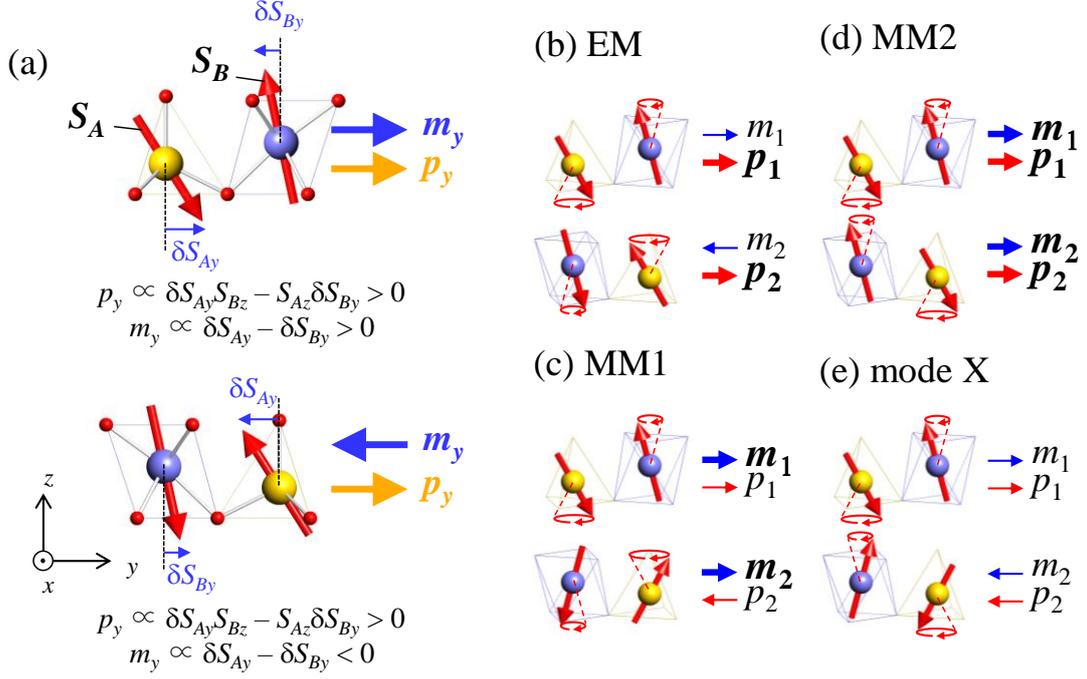

Fig. 4

(a) Schematic dynamical spin configurations with oscillating in-plane magnetization ($m_y \propto \delta S_{Ay} - \delta S_{By}$) and electric polarization ($p_y \propto \delta S_{Ay}S_{Bz} - S_{Az}\delta S_{By}$) for upper and bottom layers in a unit cell. (b)-(e) Corresponding spin configurations for the respective magnetic excitation modes.

Acknowledgement

The authors thank Y. Okamura, L. Ye, for enlightening discussions. This work was partly supported by Grants-In-Aid for Scientific Research (Grant No. 24224009) from the MEXT of Japan.